\documentclass[aps,prb,onecolumn,showpacs,showkeys,floatfix,groupedaddress]{revtex4-2}

\usepackage{amsfonts,fancybox,pgf}
\usepackage{amssymb}
\usepackage{float}
\usepackage{footnote}
	\usepackage{amsmath}%,amssymb}
	\usepackage{makeidx}
	\usepackage{amsfonts}
	\usepackage[ansinew]{inputenc}
	\usepackage[usenames,dvipsnames]{pstricks}
	\usepackage{subfigure}
	\usepackage{graphicx}
	\usepackage{here}
    \usepackage{epstopdf}
	\usepackage{epsfig}
	\usepackage{pst-grad} % For gradients
	\usepackage{pst-plot} % For axes
	\usepackage[colorlinks,hyperindex]{hyperref}
	\usepackage[english]{babel}
	\hypersetup
		{
		colorlinks,%
		citecolor=blue,%
		linkcolor=black,%
		urlcolor=black,%
	}

\makeindex

\begin{document}

\title{\normalsize Induced phase transitions and spontaneous symmetry breaking based on the renormalized Ginzburg-Landau theory\\}
%\title{\normalsize \textcolor{red}{Non-universal} Phase transitions and spontaneous symmetry breaking in the renormalized Ginzburg-Landau theory}

\author{O. C. Feulefack$^{a}$, C. Tsague Fotio$^{b}$, R. M. Keumo Tsiaze$^{a,c}$, S.E. Mkam Tchouobiap$^{a,d}$, M.N. Hounkonnou$^{a}$}

\address{$^{a}$ International Chair in Mathematical Physics and Applications, (ICMPA-UNESCO Chair), University of Abomey-Calavi, 072 P O Box 50, Cotonou, Republic of Benin}
\address{$^{b}$ Laboratory of Condensed Matter-Electronics and Signal Processing, Department of Physics, Faculty of Science, University of Dschang, P.O. Box 479 Dschang, Cameroon.}
\address{$^{c}$ Laboratory of Mechanics, Materials and Structures, Faculty of Science, University of Yaound\'{e} I, P.O. Box 812, Yaound\'{e}, Cameroon}
\address{$^{d}$ Laboratory of Research on Advanced Materials and Nonlinear Science (LaRAMaNS), Department of Physics, Faculty of Sciences, University of
Buea, PO Box 63, Buea, Cameroon}
\email{ornelafeulefack@gmail.com, carlostsague@gmail.com, keumoroger@yahoo.fr, esmkam@yahoo.com, hounkonnou@yahoo.fr}
%\vspace{10pt}
%\begin{indented}
%\item[]May 2021
%\end{indented}

\begin{abstract}
	
In this study, we present theoretical investigations of phase transitions and critical phenomena in materials through the lens of second-order Ginzburg-Landau theory, in conjunction with considerations of symmetry groups and thermal fluctuations. By addressing the residual effects after a renormalization process, a small number of macroscopic degrees of freedom can effectively replace the infinite number of microscopic degrees of freedom, emphasizing the significant role of dimensionality and the intrinsic characteristics of the system in understanding and analyzing transitions. We highlight several non-universal characteristics of continuous phase transitions near the transition temperature, including the non-monotonic relationship between the critical temperature and dimensionality, as well as the enhancement or disappearance of the specific heat jump in complex superconductors. While the resulting expressions for thermodynamic quantities are complex for one-dimensional systems, obeying Mermin-Wagner's theorem, they are considerably simplified for two-dimensional and three-dimensional systems.
\end{abstract}

% Uncomment for keywords
\vspace{0.5pc}


% Uncomment for Submitted to journal title message
%\submitto{{\color{blue}.....}}
%
% Uncomment if a separate title page is required
\maketitle
%
% For two-column output uncomment the next line and choose [10pt] rather than [12pt] in the \documentclass declaration
%\ioptwocol

\section{Introduction}\label{Sec1}
\noindent

So far, the miniaturization of devices has led to several discoveries and the creation of new functions resulting from the microscopic properties/characteristics of materials. For instance, a higher density of materials in a single direction or plane is made possible by reducing the thickness of the bulk material, which also permits the downsizing of other dimensions. Accordingly, understanding and designing new materials with different geometries and symmetries, which generally result in new functions and demonstrate many intriguing properties, i.e., tremendous characteristics from the standpoint of phase transitions, continue to considerably draw the attention of many researchers~\cite{Landau1,HohenbergPC}.  

Symmetry breaking in physics refers to a phenomenon where a symmetric and disordered state collapses into an ordered but less symmetric state~\cite{Landau2,Ginzburg,HohenbergPC}. In certain cases, this transition represents a bifurcation that a particle can undergo as it moves toward a lower energy state, as illustrated in Fig. 1. When considering non-abelian aspects, symmetry breaking results in degeneracy in the spectrum, since the states in the theory furnish a representation of the symmetry. Amongst other things, spontaneous symmetry breaking is well known as a key concept in statistical field theory and plays a crucial role in the so-called Ginzburg-Landau theory of phase transitions and critical phenomena~\cite{Landau1,Landau2,Ginzburg,Stanley,HohenbergPC,Privman}.
\begin{figure}
\begin{center}
\includegraphics[width=12cm]{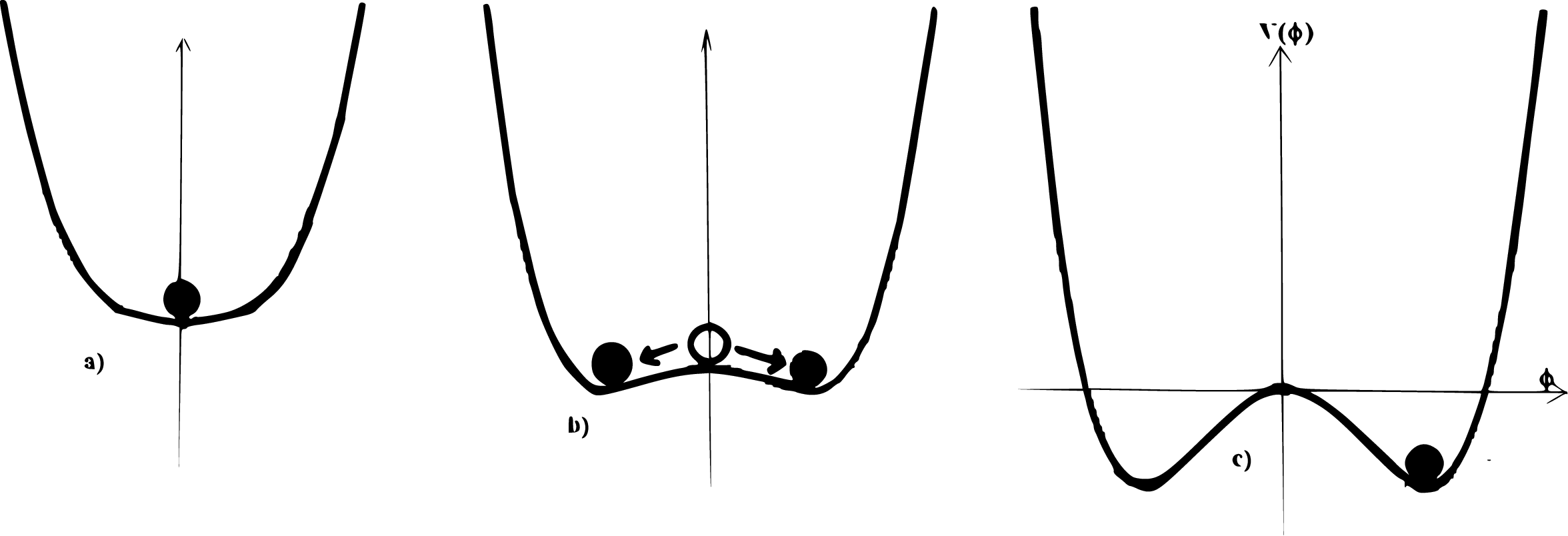}
\caption{Schematic illustrations of the evolution of the effective on-site potential corresponding to: a) non-breaking/Unbroken symmetry, b) proposed ${\mathbb{Z}_{2}}$-symmetric system with two possible states, and c) effective symmetry breaking/ Broken symmetry. This general framework identifies the mechanisms and provides the guideline for the phase transition mechanism.}
\end{center}
\end{figure}

More importantly, symmetry breaking arguments in general are powerful tools to deduce and control critical qualitative properties of phase transitions, further leading to a striking pattern with lower symmetry, i.e., breaking the original permutation symmetries of unit cells~\cite{Landau1,Hohenberg1} and also to macroscopic buckling states~\cite{Wu,Kang}. Indeed, the Landau model assumes that a thermodynamic potential $H_{L}[\phi]$ can be expressed as a series in terms of the order parameter $\phi$, and the model analyzes the symmetry properties of different phases and has been extensively studied, as referenced in the review by Hohenberg and Krekhov~\cite{Hohenberg1}. In this paper, we will focus on transitions characterized by order parameters where the symmetry group of the least symmetric phase (the ordered phase) is a subgroup of the symmetry group of the most symmetric phase (the disordered phase). Specifically, we will examine cases where the order parameter is continuous, leading to second-order transitions. In these situations, the free energy  $H_{L}[\phi]$ is analytic in terms of $\phi$ and temperature for a spatially uniform system. Near the transition temperature, this free energy can be expressed in a specific form,
\begin{equation}
H_{L}[\phi] = V\Big(\frac{r}{2}\phi^{2} + \frac{b}{4}\phi^{4}  - h\phi\Big).
\label{eq1}
\end{equation}
In this context, $V$ represents the volume of the system, while $h$ denotes an external field. The equilibrium value of the order parameter is determined by minimizing the Landau functional.
\begin{equation}
\frac{\partial{H_{L}[\phi]}}{\partial\phi} = 0 \phantom{..} \Rightarrow \phantom{..} r\phi + b\phi^{3} = h.
\label{eq2}
\end{equation}
Importantly, in zero field, two scenarios can arise, which can be summarized as follows:

$\bullet$ The value of \(\phi = 0\) represents a minimum of the potential and indicates a stable equilibrium position (see Fig. 1(a)). The concept of spontaneous symmetry breaking primarily revolves around the sign of the quadratic term \(r\) in \(H_{L}[\phi]\). In accordance with Landau's foundational framework for phase transitions in the disordered phase at high temperatures, \(\phi = 0\) is the absolute minimum of \(H_{L}[\phi]\) and corresponds to a point that is invariant under the symmetry. In this scenario, where \(r > 0\), we refer to the symmetry as being unbroken, as demonstrated in Fig. 2 for temperatures \(T > T_c\), where \(T_c\) denotes the mean-field transition temperature, not accounting for fluctuation effects.

$\bullet$ ${\phi = 0}$ represents a relative maximum, leading to two other stable equilibrium positions, ${\phi}_{\pm} = \pm{{\phi_0}}$, as illustrated in Figs. 1(b) and 1(c). When $r < 0$, the $\phi^2$ term in Eq.~\eqref{eq1} has a negative quadratic coefficient, making $\phi = 0$ a local maximum instead of a minimum. Below the transition temperature $T_c$, but close to it, the potential reaches an absolute minimum near $\phi = 0$. In this ordered (low-temperature) phase, the system can adopt either of the two degenerate values $\phi_{\pm}= \pm|\phi_{0}|$, which necessitates $r < 0$ and  $b > 0$, as shown in Fig. 2 for temperatures $T < T_c$. In fact, $r > 0$ results in a discontinuous transition, where $\phi$ shifts from zero to a finite value at $T_c$. The same consequence follows for $b < 0$. Since $r$ changes sign at $T_c$ and regularity has been assumed for all thermodynamic variables, Landau's basic construction leads to the expression for the quadratic coefficient: $r = r_{0}(T - T_{c})$, where $r_{0}$ is a constant~\cite{Landau1,Stanley,Hohenberg1}.
\begin{figure}
\begin{center}
\includegraphics[width=12cm]{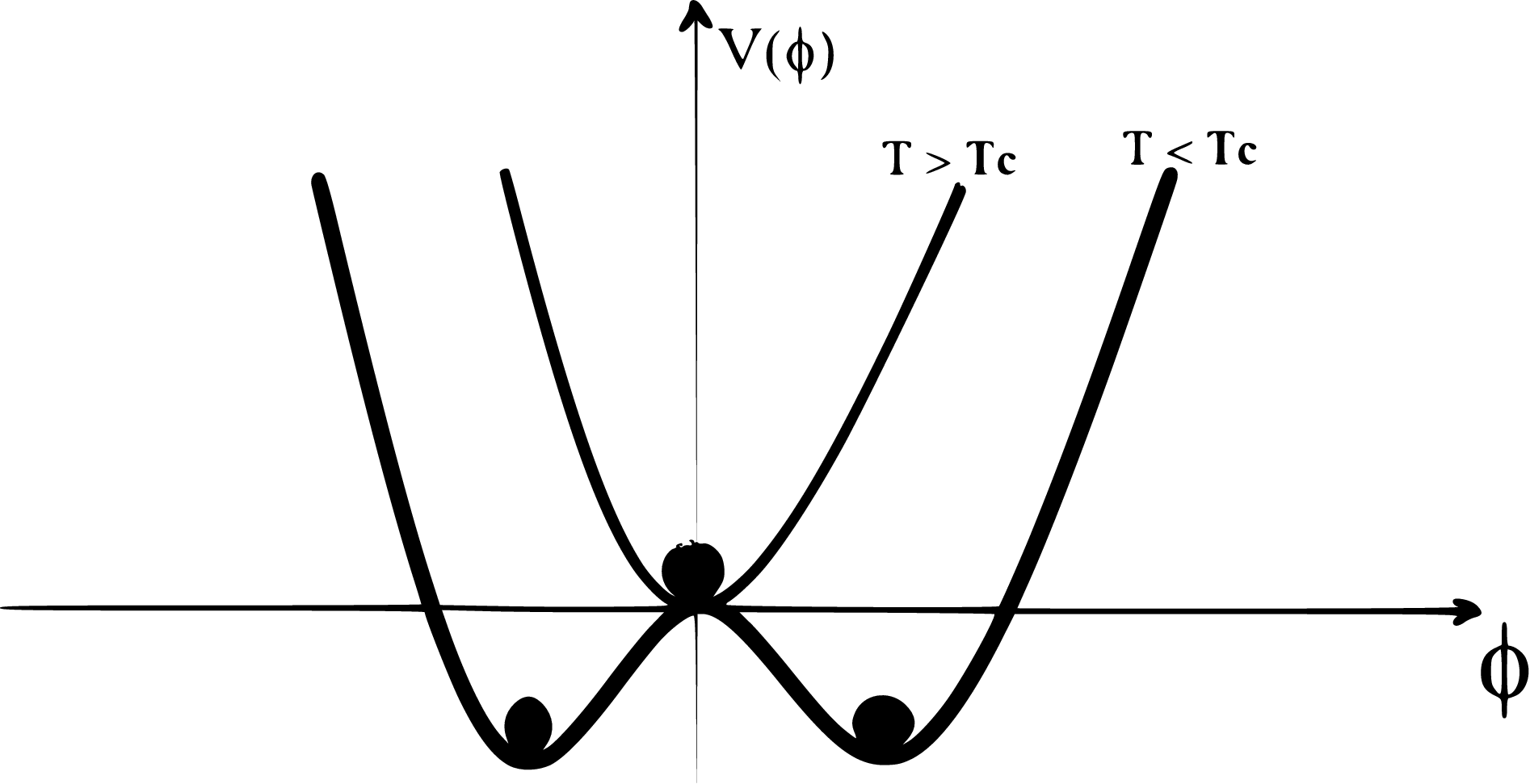}
\caption{Schematic geometry representation of the evolution of the effective on-site potential with temperature, illustrating spontaneous symmetry breaking behavior at $T_{c}$ and the structural phase transition mechanism of the multistable systems.}
\end{center}
\end{figure}

In general, the minimization of the Landau functional without taking into account an external field leads to:
\begin{eqnarray}
&&\phi_0^2 =  \left\{  \begin{array}{ll}\vspace{0,5cm}
0, & \textrm{for}\phantom{.} \tau > 0 \phantom{...} \textrm{characterizing the disordered phase}  \\
-r/b, & \textrm{for} \phantom{.} \tau < 0 \phantom{...} \textrm{characterizing the ordered phase}. \\
  \end{array}\right.
  \label{eq3}
\end{eqnarray}
Here, $\tau = \frac{T - T_c}{T_c}$ is the reduced temperature. The specific heat $C_p$ is expressed by substituting the equilibrium value of the order parameter into the free energy given in Eq.~\eqref{eq1}. This expression can be factorized as follows:
\begin{eqnarray}
&&C_{p} =  \left\{  \begin{array}{ll}\vspace{0,5cm}
C_0, & \textrm{for}\phantom{...} \tau > 0 \\
C_0 + \frac{r_0^2T}{2b}, & \textrm{for} \phantom{...} \tau < 0 \\
  \end{array}\right.
  \label{eq4}
\end{eqnarray}
In this context, $C_0$ represents a specific heat constant, while $\Delta C_0 = \frac{r_0^2T}{4b}$ indicates the mean-field jump in the specific heat as the temperature falls below the transition temperature $T_c$. This specific heat jump varies in prominence depending on the intrinsic properties of certain materials, and it may even vanish entirely in some complex superconducting systems. This review will focus on the phenomenon of the disappearance of the specific heat jump, with the aim of identifying its underlying causes.

Reconsidering an external field and utilizing Eq.~\eqref{eq3}, differentiating Eq.~\eqref{eq2} provides the expression for susceptibility as follows:
\begin{eqnarray}
&&\chi^{-1} = \frac{\partial{h}}{\partial\phi} = r + 3b\phi^{2} \phantom{..} \Rightarrow \phantom{.} \chi^{-1}\Big\vert_{h\rightarrow 0} =  \left\{  \begin{array}{ll}\vspace{0,5cm}
r, & \textrm{for}\phantom{.} \tau > 0 \\
-2r, & \textrm{for} \phantom{.} \tau < 0. \\
  \end{array}\right.
  \label{eq5}
\end{eqnarray}
A positive aspect of the model $H_{L}$ is that it predicts a transition at a non-zero critical temperature, $T_c$, which aligns with experimentally observed properties. The order parameter $\phi_{h = 0}$ is non-zero at temperatures below $T_c$. Additionally, the susceptibility at zero external field, $\chi_{h = 0}(T)$, diverges at the critical temperature, $T_c$. However, a negative aspect is that the values of the critical exponents derived from the model do not match the experimental results. While Landau theory is qualitatively valid, it is quantitatively inaccurate.

There are significant size effects in low-dimensional systems that are negligible in bulk materials. Therefore, the interest of this work lies in understanding and designing materials with various geometries and symmetries from the perspective of structural phase transitions, and which induce orderly localized configurations.
Moreover, as well known in the phase transition theory considered up to now, the spontaneous symmetry breaking came from the phase transition; so we will focus on systems with spontaneous symmetry breaking, so that Ginzburg-Landau theory will once again turn out to be useful. In this context and to be more specific, we will rely on a consideration of a system undergoing a continuous phase transition from a high-temperature symmetric phase to a low-temperature ordered phase in which some symmetry is broken.

In this review, we develop and summarize ideas that lead to a theoretical model of low-dimensional systems. The outline of the review is organized as follows: In Section~\ref{Sec2}, we present the second-order Ginzburg-Landau theory, highlighting its limitations and the necessity of making corrections through a renormalization process. This process emphasizes the critical role of dimensionality in the manifestation and analysis of transitions. Our theoretical results, computed using typical parameters, demonstrate the importance of considering both dimensionality and thermal fluctuations, confirming that size effects significantly influence the thermodynamic properties of materials. Unlike the standard Landau model, this review aims to provide a corrective term for the quadratic coefficient, incorporating contributions from symmetries and thermal fluctuations. We derive an appropriate expression for this quantity in terms of the system's correlation functions and scattering state contributions. The spatial variations of the order parameter in a fluctuating system are described by a nonlinear polynomial equation, which takes into account the system's anisotropy and the dependence of the quadratic coefficient on the intrinsic characteristics of the system. By averaging over the system's fluctuations, we obtain a macroscopic equation for spatial variations that features a dimension-dependent quadratic coefficient. We derive alternative exact expressions for the renormalized quadratic coefficients that contain more information. In Section~\ref{Sec3}, we highlight evidence that the obtained model allows for better simulation of certain physical conditions involved in phase transitions and predicts specific physical constraints. This is achieved by re-expressing key thermodynamic parameters, notably the jump in specific heat and susceptibility, both of which depend on the system's dimensionality. Accordingly, Section~\ref{Sec4} presents, we provide a comprehensive discussion of the results and their implications. Finally, Section~\ref{Sec5} offers a conclusion.

 \section{The renormalized Ginzburg-Landau theory}\label{Sec2}
\noindent

The basic reason for the failure of Landau's theory is the neglect of spatial fluctuations, which stems from the assumption that the order parameter is spatially homogeneous. To account for the effects of these spatial fluctuations, the order parameter should be treated as a tensor field instead of just a constant. Consequently, the Landau potential is replaced by an integral over a potential free energy density, known as the Ginzburg-Landau Hamiltonian functional \cite{Ginzburg},
\begin{equation}
H_{GL}[\phi] = \int d^{d}\textbf{r}\bigg[\frac{\mathcal{C}}{2}\big(\nabla_r\phi(\textbf{r})\big)^2 + \frac{r}{2}\phi^{2} + \frac{b}{4}\phi^{4} - h\phi \bigg].
\label{eq6}
\end{equation}
Here, the coefficients $\mathcal{C}$ and $b$ are assumed to be weakly dependent on external thermodynamic parameters (e.g. temperature and pressure).

Although Ginzburg-Landau theory accounts for fluctuations, it has limitations when it comes to describing systems with reduced dimensionality. For example, it cannot adequately explain the order of phase transitions in certain layered bulk materials~\cite{Cybart}. Additionally, the theory falls short in addressing one-dimensional~\cite{Scalapino} and quasi-two-dimensional systems, where new states with unexpected properties emerge at non-zero temperatures~\cite{Tinkham}. Another significant drawback of Ginzburg-Landau theory is its inability to account for the absence or disappearance of the specific heat jump observed in some materials, such as YBa$_2$Cu$_3$O$_{6 +\delta}$~\cite{Loram}, Ba$_{0.2}$K$_{0.8}$Fe$_2$As$_2$~\cite{Tanaka} and certain families of high-$T_c$ superconductors, such as TlBaCaCuO and BiSrCaCuO~\cite{Meingast}.

\subsection{Propositions}\label{Sec2.1}
\noindent
The importance of critical fluctuation effects, their dynamics, and their impact on the physical properties of materials has been explored in various studies~\cite{Anatoly,Doniach,Zinn-Justin,Ma,Amit, Papon,Varlamov,Kleinert}. However, there are still common challenges when it comes to assigning the parameters of the standard Ginzburg-Landau theory to their characteristic quantities, such as temperatures, in a universal way~\cite{Poole}. The renormalization process refines the standard Ginzburg-Landau theory by adding a correction term to the quadratic coefficient, which isn't accounted for in the standard model. By following this approach, effects of fluctuations and correlations of order parameter are incorporated by replacing the quadratic coefficient with its dimension-dependent expression
\begin{equation}
r_0(T - T_{c}) \rightarrow r_0(T - T_{c}) + \Omega(d, T).
\label{eq7}
\end{equation}
By employing this transformation, we can frame our discussion by qualitatively comparing the results obtained for various dimension values. The model derived from the transformation described in Eq. (7) is phenomenological, as we propose the existence of fluctuations without detailing their origins. In this context, the auto-coherent dimension-dependent correction quantity $\Omega(d, T)$, first derived within the framework of the renormalized Gaussian approach by Keumo \textit{et al.} and which satisfy the following self-consistent equation (see appendix~\ref{App1})~\cite{Keumo1}:
\begin{equation}
 \Omega(d, T)=\eta\left(\frac{T}{T_{c}}\right)\left[\tau +\frac{\Omega(d, T)}{r_{0}T_{c}}\right]^{\frac{d}{2}-1}, \phantom{...} \textrm{where} \phantom{...} \tau  = \frac{T - T_{c}}{T_{c}},
 \label{eq8}
 \end{equation}
contains important information about regulating the anisotropy of the order parameter. 
This expression shall for historical reasons referred to as 'Keumo-Mkam-Danga-Domngang-Hounkonnou'(KMDDH) correction coefficient relation of the quadratic expansion of the Ginzburg-Landau free energy functional induced by the quantum fluctuations of the order parameter.

Here the coefficient $\eta$ denotes a variational energetic parameter related to the harmonic variance of the order parameter which takes into account the proper dependence of the magnitude and the temperature of the fluctuation corrections on the anisotropy of the electron spectrum and the possibly related anomalies including the impurity concentration, stacking faults and other intrinsic defects shown by physical properties~\cite{Larkin,Varlamov,Keumo2}. More specifically, the quantity $\eta$ generally depends on both the coherence length and the fourth order coupling constant which is known to account for the self-interaction of the system and the control of the saturation of ordering in the Ginzburg-Landau theory, and its magnitude can induce or prevent phase transitions~\cite{Keumo2}. Furthermore, the Ginzburg-Landau Hamiltonian $H_{GL}$, as a function of a spatially varying generalized single order parameter configuration $\phi(\textbf{r})$ is now renormalized, which finally leads us to a coarse-grained effective Ginzburg-Landau Hamiltonian involving fluctuations and in the presence of an external field h, and which may now cast the form
\begin{equation}
H_{RGL}[\phi] = \int d^{d}\textbf{r}\bigg[\frac{\mathcal{C}}{2}\big[\nabla_r\phi(\textbf{r})\big]^2 + \frac{r(d,T)}{2}\phi^{2} + \frac{b}{4}\phi^{4} - h\phi \bigg].
\label{eq8bis}
\end{equation}
where the phenomenological renormalized quadratic coefficient $r(d, T) = r_0(T - T_{c}) + \Omega(d, T)$. This new coefficient takes into account the fluctuations on the two-point correlation function, which implies the destruction of the long-range order, as predicted by the mean-field theory~\cite{Landau1,Ginzburg,Stanley,Hohenberg1}.

By competing with $r_0(T - T_c)$ in the renormalized Ginzburg-Landau Hamiltonian Eq.~\eqref{eq8bis}, the dimension-dependent function $\Omega(d, T)$ helps eliminate divergences in field theory~\cite{Tsague}. We conceptualize the renormalization process as a collection of techniques designed to address the infinities that arise in calculated thermodynamic quantities. This is achieved by modifying expressions of these quantities to account for the effects of their self-interactions. Using a self-consistent approximation for the quadratic coefficient, we aim to achieve modifications in thermodynamic parameters that incorporate fluctuations, thereby removing the unphysical maximum observed in the crossover regime of standard theory. The inclusion of the $\eta$ term accounts for contributions from scattering states and repulsive interactions between quasi-particles. This competition between $r_0(T - T_c)$ and $\Omega(d, T)$ suggests the coexistence of competing phases within the system, with each phase corresponding to a spontaneously broken symmetry. As a result, the repulsive fluctuation effects cause  non-synchronized behavior among quasi-particles, significantly altering the underlying physics~\cite{Keumo2}. This concept of competing states helps explain the emergence of new states that remain unexplained, depending on whether attractive or repulsive interactions dominate. Fluctuation effects, similar to the orbital effect, have a greater impact on lower phase destruction compared to higher phases. Considering the contributions from scattering states at the Gaussian level clarifies why, in standard Ginzburg-Landau theory, both the dimensional and temperature dependencies of the quadratic coefficient must be taken into account. Even when the paramagnetic effect becomes predominant, this approximation holds true. We will further discuss the significance and role of the energy parameter $\eta$ later. 

\subsection{Dimension-dependent renormalized quadratic coefficient and treatment}\label{Sec2.2}
\noindent 
In order to illuminate the physics of thermodynamic phases and leads to powerful theoretical methods for understanding critical phenomena at continuous phase transitions, it is important to investigate  quite generally the dependence of the critical point on the quantum fluctuations throughout the scaling and fluctuation corrections. In this context, the primary objective of the present study is to investigate the effects of spatial variations and fluctuations on the Ginzburg-Landau functional expansion.
Accordingly, to calculate thermodynamic quantities directly from the renormalized Ginzburg-Landau functional, we need to identify the true critical temperature, incorporating the correction term. This critical temperature will be denoted as $T_{c_d}$, where the index $d$ indicates the dimension of the system. By using the expression for the KMDDH correction term $\Omega(d, T)$ from Eq.~\eqref{eq8}, we can derive an equation for the dimension-dependent renormalized quadratic coefficient $r(d, T)$ given by the following self-consistent dimension-dependent KMDDH polynomial equation~\cite{Keumo1}:
\begin{equation}
r(d, T) - \eta\left(\frac{T}{T_{c}}\right)\Bigg[\frac{r(d, T)}{r_0T_c}\Bigg]^{d/2 - 1} - r_0(T - T_c) = 0.
\label{eq9}
\end{equation}

Upon examining the important result presented in Eq.~\eqref{eq9} more closely, we observe that fluctuations increase with $r(d, T)$ raised to the power of $d/2 - 1$, as long as fluctuation corrections cannot be ignored for dimensions $d \leq 2$. $d_c$ = 2 represents the critical dimension for this approach, below which fluctuations become highly significant. This observation aligns with the Hohenberg-Mermin-Wagner theorem, which states that continuous symmetries cannot be spontaneously broken at finite temperatures in systems with sufficiently short-range interactions when the dimension is $d \leq 2$~\cite{Mermin,Hohenberg}.

By considering the different physical dimension $d$, solutions of Eq.~\eqref{eq9} is dimension-dependents and are expressed by the following renormalized quadratic coefficients,
\begin{eqnarray}
&&r(d, T) =  \left\{ \begin{array}{llll}\vspace{0,5cm}
r(d = 1, T) & \textrm{for}\phantom{.} d = 1  \\ \vspace{0,3cm}
\eta\left(\frac{T}{T_{c}}\right) + r_0(T - T_{c}), & \textrm{for} \phantom{.} d = 2,\\ \vspace{0,3cm}
\frac{\eta{T}\Big(\eta{T} \pm\sqrt{\eta^2T^2 + 4r^2_0T^3_{c}(T - T_{c})}\Big)}{2r_0T^3_{c}} + r_0(T - T_{c}), & \textrm{for} \phantom{.} d = 3,\\ \vspace{0,3cm}
r_0\big(1 - \frac{\eta{T}}{r_0T^2_{c}}\big)^{-1}(T - T_{c}), & \textrm{for} \phantom{.} d = 4.\\
  \end{array}\right.
 \label{eq10}
\end{eqnarray}

Now, let us consider the special case of one dimension ($d$ = 1) related to the following equation:
\begin{equation}
r(d = 1, T) - \eta\left(\frac{T}{T_{c}}\right)\sqrt{\frac{r_0T_c}{r(d =1, T)}} - r_0(T - T_c) = 0.
\label{eq11}
\end{equation}
The equation presented in Eq.~\eqref{eq11} outlines a self-consistent condition related to the renormalized quadratic coefficient $r(d = 1, T)$. A detailed mathematical analysis of this equation reveals that $r(d = 1, T)$ is complex and consistently positive, regardless of the temperature. It is computed using typical numerical parameters and is represented by the dashed curve in Fig. 3, illustrating its behavior. At very low temperatures, $r(d = 1, T)$ exhibits temperature-independent behavior and asymptotically approaches zero. As a result, for $d$ = 1, symmetry cannot be broken, indicating that $\phi = 0$ is the absolute minimum of the renormalized potential $H_{L}[\phi] = \frac{r(d = 1, T)}{2}\phi^2 + \frac{b}{4}\phi^4$. This corresponds to a point that is invariant under the symmetry. To explore the new contributions, we make a qualitative comparison between the results obtained for standard and renormalized models, focusing primarily on the quantities $r(d = 1, T)$ and $r_0(T - T_c)$ following the contribution of the correction term $\Omega_{1D}(T)$. The results of this comparison are also shown in Fig. 3. From this analysis, we can conclude that long-range order due to the breaking of continuous symmetry is absent in the renormalized Landau theory for one-dimensional systems.

\begin{figure}
\begin{center}
\includegraphics[width=12cm]{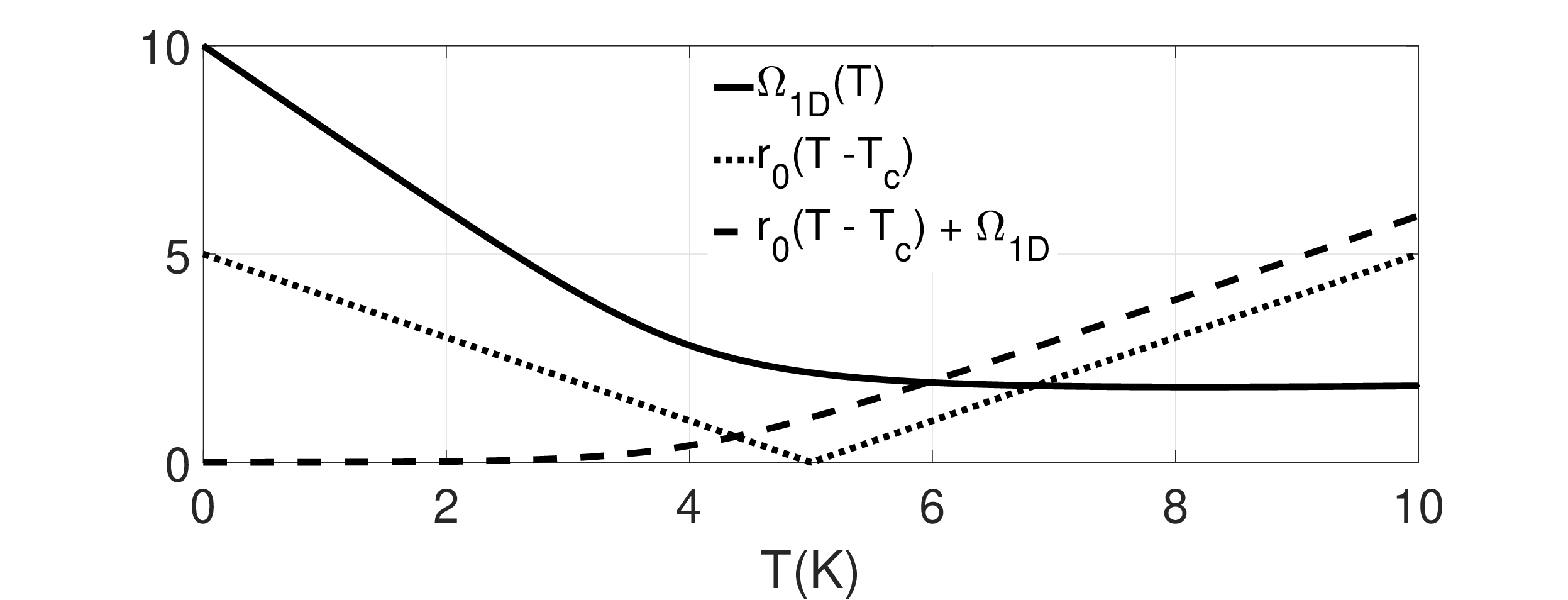}
\caption{The plot illustrates the standard and renormalized quadratic coefficients for a one-dimensional system. It is assumed that $T_{c} = 5 K$ and  $\eta$ = 0.8. The solid curve represents the correction term $\Omega_{1D}(T)$, the dotted curve shows the standard quadratic coefficient $r_0(T - T_c)$, and the dashed curve depicts the behavior of the renormalized quadratic coefficient, defined as $r(d, T) = r_0(T - T_c) + \Omega_{1D}(T)$. While the standard coefficient goes to zero at $T_c$, the renormalized quadratic coefficient asymptotically approaches zero, i.e. $T_{c_{1D}}$ = 0 is the critical temperature for one-dimensional systems in the renormalized theory.}
\end{center}
\end{figure}

It is important to precise that, the obtained renormalized quadratic coefficients still cancel out at a finite transition temperature given by $r(d, T)$ = 0 (except for one-dimensional systems), but the critical behavior is qualitatively and quantitatively modified. However, the transition temperature is no longer monotonic because it depends on the system's dimensionality.
\begin{eqnarray}
&&r(d, T) = 0 \Rightarrow  \left\{ \begin{array}{llll}\vspace{0,3cm}
T_{c_{1D}} = 0, & \textrm{for}\phantom{.} d = 1  \\ \vspace{0,3cm}
T_{c_{2D}} = T_c(1 + \epsilon)^{-1}, & \textrm{for} \phantom{.} d = 2,\\ \vspace{0,3cm}
T_{c_{3D}} = T_c, & \textrm{for} \phantom{.} d = 3,\\ \vspace{0,3cm}
T_{c_{4D}} = T_c , & \textrm{for} \phantom{.} d = 4,\\
  \end{array}\right.
  \label{eq12}
\end{eqnarray}
where $\epsilon = \frac{\eta}{r_0T_{c}}$ defines the scaled fluctuation correction ratio term.
For $d$ = 2, the expression for the renormalized quadratic coefficient is given by $r(d = 2, T) = \eta\left(\frac{T}{T_{c}}\right) + r_0(T - T_c)$. This expression is relatively straightforward, similar to the thermodynamic quantities derived from it. The temperature at which $r(d = 2, T)$ changes sign decreases as $r(d = 2, T)$ increases and is significantly influenced by the nature of fluctuations in the system (see Eq.~\eqref{eq12} for $d$ = 2)). For large values of $\eta$, $T_{c_{2D}}$ can approach zero, thereby confirming the Mermin-Wagner theorem~\cite{Mermin}. In the following paragraph, we will discuss in detail the behavior of the transition temperature in two-dimensional systems. For $d > 2$, the solutions are nearly the same as those obtained from standard theory, with the main difference being that thermodynamic parameters depend on the intrinsic nature of the systems involved. While Landau's theory does not allow us to predict the exact transition temperature, it does facilitate an analysis of the system's behavior near the transition. In our model, we emphasize that the mean-field transition temperature represents a characteristic thermal scale rather than the true transition temperature for systems with dimensions $d \leq 2$. This highlights the necessity of modifying Landau's theory to account for the crucial role of dimensionality in the manifestation and analysis of transitions, which is important for understanding concrete phenomena. Several open questions in phase transitions and critical phenomena remain pertinent. A key question is whether all low-temperature phase effects arise solely from fluctuation-induced order parameters. Given that pre-transitional behaviors provide insights into the physical mechanisms underlying phase transitions, our approach diverges from the standard one and necessitates numerical investigation of Eqs.~\eqref{eq8} and ~\eqref{eq9} to better understand the contribution of the correction term $\Omega(d, T)$.

\section{General framework and Renormalized thermodynamic quantities}\label{Sec3}
\noindent
Having determine the fluctuations contribution at quadratic order to the free energy, it is necessary to evaluate the fluctuation contribution to some thermodynamic quantities.
As well-known, the universality that characterizes the behavior of thermodynamic functions close to a second order phase transition appears as one of the fascinating aspects of critical phenomena in statistical mechanics~\cite{Stanley,Ma,Cardy}.
By applying Fourier modes as presented in the appendix in the Ginzburg-Landau functional and in the absence of an external field, we can establish a proportional relationship between the variance of the order parameter (which represents fluctuations) and the average change of that order parameter when the external field changes. This relationship, known as the fluctuation-dissipation theorem, can be expressed as follows:
\begin{equation}
\chi = \frac{1}{T}\int d^{d}\textbf{r}C(\textbf{r}) \phantom{..} \Rightarrow \phantom{..} \chi(\textbf{q}) = \frac{V}{T}C(\textbf{q}).
\label{eq18}
\end{equation}
which relates the susceptibility $\chi$ and the correlation function $C$. More generally, this fluctuation-response relation is valid for weak fluctuations since linear response is assumed.
Importantly, in the absence of an external field ($h = 0$) and referring to Eq.~\eqref{eq5}, one can obtain the susceptibility modified by fluctuations in the following manner:
\begin{eqnarray}
\chi^{-1}\Big\vert_{h\rightarrow 0} =  \left\{  \begin{array}{ll}\vspace{0,5cm}
r_0(T - T_{c}) + \Omega(d, T), & \textrm{for}\phantom{.} T > T_{c_d}, \\
2\Big(r_0(T_{c} - T) - \Omega(d, T)\Big). & \textrm{for} \phantom{.} T < T_{c_d}.\\
  \end{array}\right.
  \label{eq19}
\end{eqnarray}
Importantly, the temperature -dependent inverse susceptibility function $\chi^{-1}(T)$ depend explicitly on the dimensional and intrinsic parameters of the system. There is a challenge when accurately evaluating certain thermodynamic quantities, such as specific heat, within the Ginzburg-Landau model. In some cases, it is necessary to apply the Gaussian approximation above the transition temperature. This approach is achieved by omitting the $|\phi|^4$-term, which is commonly linked to the neglect of scattering state contributions in Eq.~\eqref{eq6}. As a result, when integrating over all fluctuation degrees of freedom, the renormalized specific heat, without considering an external field, can be factorized as follows:
\begin{eqnarray}
&&C_p =  \left\{ \begin{array}{ll}\vspace{0,5cm}
\frac{1}{2}\gamma_d{(\xi^{+}_0)^{-d}}\Big[\frac{T}{T_c}\Big]^2\Big(1 + T_c\frac{\partial\kappa(d, T)}{\partial T}\Big)^2\tau_d^{-\alpha}, & \textrm{for}\phantom{...} T < T_{c_d}, \\
\Delta{C_0}(d, T_{c_{d}}) +   \gamma_d{(\xi^{-}_0)^{-d}} \Big[\frac{T}{T_c}\Big]^2\Big(1 + T_c\frac{\partial\kappa(d, T)}{\partial T}\Big)^2\Big\vert\tau_d\Big\vert^{-\alpha'}, & \textrm{for} \phantom{...} T < T_{c_d} ,\\
  \end{array}\right.
  \label{eq20}
\end{eqnarray}
where $\tau_d = \tau + \kappa(d, T)$ is the renormalized reduced temperature. The (Gaussian) critical exponents $\alpha = \alpha' = 2 - d\nu$ with $\nu$ = 1/2. $\xi^{\pm}$ is the coherence length of th system above (+) and below (-) the renormalized critical point. The dimensional constant $\gamma_d$ is an integral quantity that arises from the Gaussian approximation and the magnitude of the ratio
\begin{equation}
\kappa(d, T) = \frac{\Omega(d, T)}{r_0T_c},
\label{eq21}
\end{equation}
determines the sequence of fluctuations in the system, which enables the transition from classical mean-field critical behavior to fluctuation-dominated critical behavior. The quantity $\kappa(d, T)$ includes scattering state contributions via the parameter $\eta$, which is essential for constructing the correction term $\Omega(d, T)$.
  
Indeed, at this extend following the more general result known as the Mermin-Wagner theorem~\cite{Mermin}, which states that there is no spontaneous breaking of a continuous symmetry in systems with short-range interactions in dimensions $d \le 2$, as the phase fluctuations become asymptotically large, leading to the corollary existence of the borderline dimensionality of two, known as the lower critical dimension $d_l$. The renormalized approach effectively illustrates and reaffirms the Mermin-Wagner theorem~\cite{Mermin}, particularly in one-dimensional systems. In two-dimensional systems, while recovering significant critical fluctuations, one observes a variation of the true critical point $T_{c_{2D}}$ between 0 K and $T_{c}$, i.e., $0 \leq T_{c_{2D}} \leq T_{c}$. 
\begin{figure}
\begin{center}
\includegraphics[width=12cm]{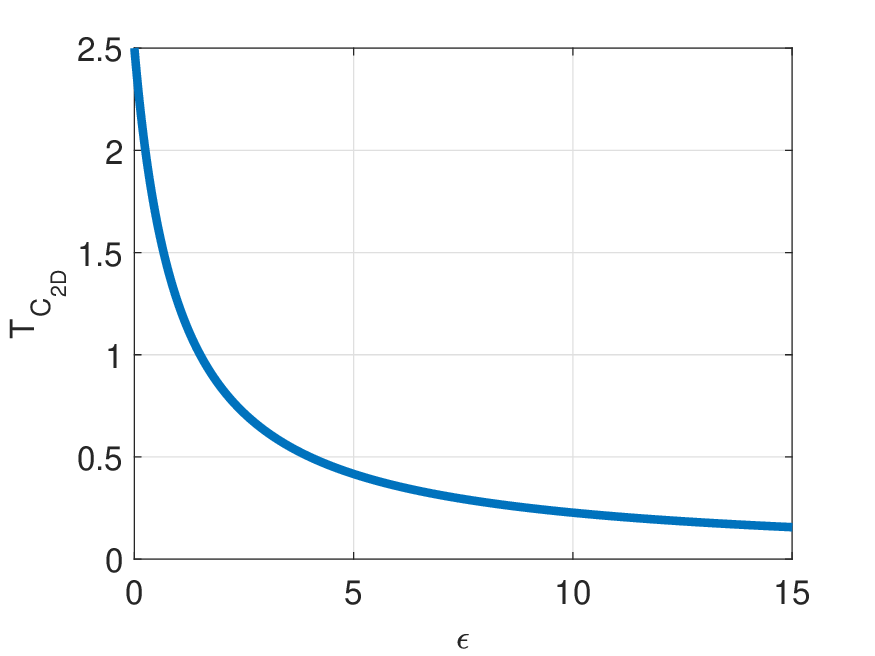}
\caption{Profile of fluctuation-correction dependence of the two-dimensional critical temperature according to Eq.~\eqref{eq12} for $T_{c}$ = 2.5 K. Here, the scaled fluctuation correction ratio $\epsilon$ defines the relative temperature distance from $T_{c}$.}
\end{center}
\end{figure}
Fig. 4 shows the temperature behavior of the true critical temperature according to the scaled fluctuation-correction term $\epsilon$. Here, a careful treatment bring a notable outcome with respect to the Mermin-Wagner theorem, which asserts that a continuous symmetry cannot be broken in two dimensions. These findings show that the presence of fluctuations leads to a decrease in the degree of order in two-dimension ($d$= 2) and completely destroys it in dimensions less than two ($d < 2$). The temperature behavior illustrated in Fig. 4 suggests a transition from classical two-dimensional fluctuations at $\epsilon = 0$ to quantum two-dimensional critical fluctuations that occur over a broad temperature range when  $\epsilon \gg 1$. For each value of the $\epsilon$ term, we can accurately determine the fluctuation term across a wide range of temperatures. Quantum fluctuations play a crucial role in driving quantum phase transitions, such as the quantum melting of Wigner crystals into Fermi liquids in electron systems. However, the effects of these fluctuations on superconducting systems close to absolute zero, especially regarding the superconductor-insulator/metal transition, remain an open question. Moreover, the parameters associated with this approach depend significantly on both the dimension and the size of the sample, but in ways that differ from those derived from the mean-field approach. As a result, the renormalized approach appears to be well-suited for model calculations in various critical systems~\cite{Keumo1,Keumo2,HohenbergPC,Keumo3}. 

From three-dimensional systems, standard and renormalization models are in good agreement, especially regarding the position of the critical point, with, however, critical behaviors being qualitatively and quantitatively different, as we will see in the following paragraph, concerning the specific heat jump $\Delta{C_p}$. However, it appears that Ginzburg-Landau models which belong to the universality class exhibit an upper critical dimension $d_u$ of four, through the so-called the Ginzburg Criterion which establishes the importance limit of fluctuations and which  has to be treated carefully. The impact of new quadratic coefficients in three dimensional systems is analyzed in greater detail elsewhere, specifically regarding the non-monotonic variation of the specific heat jump, $\Delta{C_p}/T_{c}$, across the transition for various high-temperature superconductors \cite{Feulefack}. For instance, in YBCO synthesized as $YBa_2Cu_3O_{7- \delta}$ with $0 \leq \delta \leq 0.18$, variations in oxygen stoichiometry can affect the density of states at the Fermi level and, consequently, the magnitude of $\Delta{C_p}$. Observations of transitions from relatively weak three-dimensional fluctuations at $\delta = 0$ to significant two-dimensional critical fluctuations can occur over a broad temperature range for $\delta \geq 0.1$ \cite{Loram3}. For each value of the phononic term $\delta$ that can be related to the term $\epsilon$, we can reliably determine the fluctuation term across a broad temperature range.

Accordingly, more specifically in Eq.~\eqref{eq10}, the four-dimensional quadratic coefficient highlights another temperature dependence, $r_4(T) \propto (T - T_{c})$, with a coefficient of proportionality that is different from $r_0$ and is also temperature-dependent. Consider a homogeneous superconductor with no superconducting current, and taking into account the minimization condition of the free energy in Eq. ~\eqref{eq8}, the critical behavior of susceptibility length and order parameter below $T_{c}$ is characterized as follows:
\begin{equation}
\chi \sim  \Big\vert{\frac{T - T_{c}}{1 - \epsilon\frac{T}{T_{c}}}}\Big\vert^{-1/2},
\end{equation}
and 
\begin{equation}
\phi \sim  \Big\vert{\frac{T_{c} - T}{1 - \epsilon\frac{T}{T_{c}}}}\Big\vert^{1/2}.
\end{equation}
Of central importance, it is noteworthy that in the function $v(T) = \Big\vert{\frac{T - T_{c}}{1 - \epsilon\frac{T}{T_{c}}}}\Big\vert$, the parameter $\epsilon$ plays a primordial role depending on whether its value is equal to, greater than or less than 1. For $\epsilon = 1$, we obtain $r_{4D}(T) = r_0$ below the critical point. In such a case, all thermodynamical quantities are constants, and the notion of fluctuation loses all its meaning in accordance with Landau's theory for $d \geq 4$. For $\epsilon > 1$, here again, we find the fundamental principles of the Ginzburg-Landau theory with an OP that increases continuously from zero starting at $T_{c}$. On the other hand, a whole new physics appears when $\epsilon < 1$, thermodynamical functions as shown in Fig. 5 and Fig. 6 can both cancel out and diverge depending on the temperature below the critical point. In physics, particularly in phase transitions that involve abrupt and discontinuous changes in the properties of systems, some functions can take the value zero at certain points and diverge, tending to infinity at other specific points. This can occur in various contexts, such as electric potentials, gravitational fields, or oscillatory phenomena. In thermodynamics, the Gibbs free energy can exhibit both cancellation and divergence at certain points that often correspond to phase transitions or critical points in the system's behavior. The cancellation of a wave function at the boundaries of a potential well is a boundary condition that ensures that the particle is confined within the well. The divergence outside the well is a mathematical consequence of the infinite nature of the potential. The cancellation and the divergence of functions in physics are often related to the nature of the forces or fields involved or to how these phenomena are modeled.

\begin{figure}
\begin{center}
\includegraphics[width=12cm]{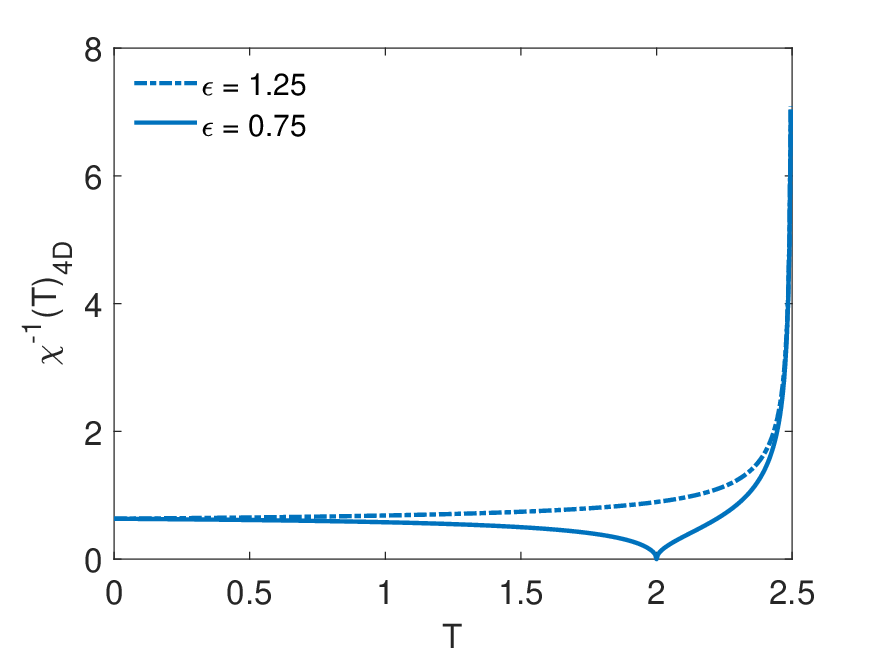}
\caption{Illustration of the susceptibility as a function of temperature in a four-dimensional system, considering various values of the scaled fluctuation-correction term, $\epsilon$. For a critical temperature of $T_{c}$ = 2.5 K, the condition  $\epsilon < 1$ indicates the relative distance between the cancellation point and the divergence point.}
\end{center}
\end{figure}

\begin{figure}
\begin{center}
\includegraphics[width=12cm]{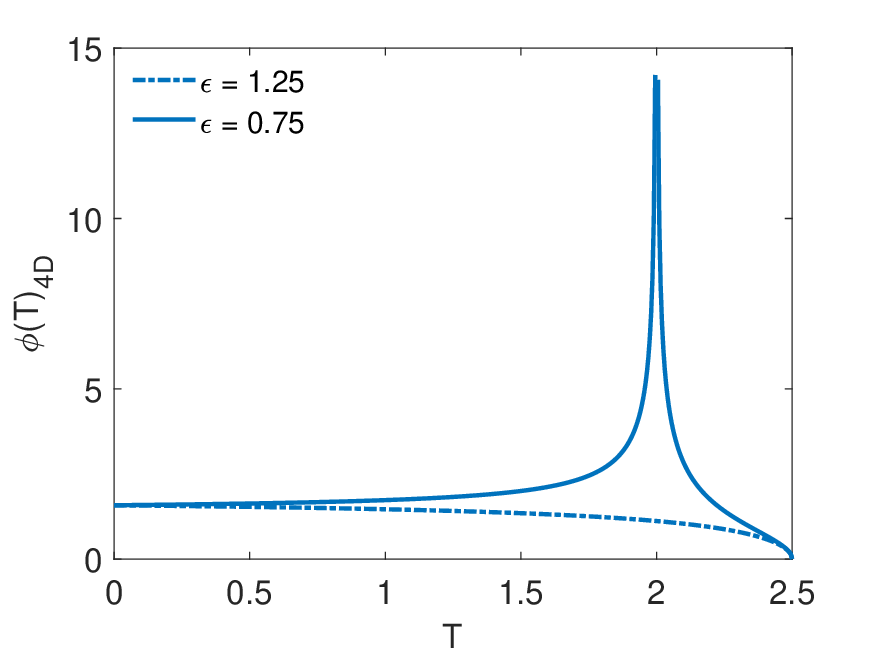}
\caption{Illustration of the order parameter as a function of temperature in a four-dimensional system, considering various values of the scaled fluctuation-correction term, $\epsilon$. For a critical temperature of  $T_{c}$ = 2.5 K, the condition $\epsilon < 1$ indicates the relative distance between the cancellation point and the divergence point.}
\end{center}
\end{figure}

Our model specifically applies to systems that have multiple competing order parameters. In certain physical systems, opposing order parameters such as electric and magnetic orders can partially cancel each other, leading to a complex overall ordering. For example, in $SU(N)$ gauge theory, a complex order parameter can be broken down into real and imaginary components. When $N \ge 3$, the various components may partially cancel each other, resulting in a complex form of ordering rather than a simpler one. A physical system that demonstrates both divergence and cancellation of an order parameter is often observed in high-energy physics. In these cases, divergences are addressed through a process known as renormalization. Similarly, in complex systems near a quantum critical point, competing order parameters can lead to cancellations. For example, in some theoretical models, divergences from one loop diagram may be canceled by divergences from other diagrams, resulting in a finite outcome. 

This specific case with $\epsilon < 1$ will be examined in greater detail elsewhere, in the context of a gravitational phase transition where the significance of dimension four is fully realized. Slow first-order phase transitions~\cite{Lewicki} can generate significant inhomogeneity, potentially leading to the formation of primordial black holes. In scenarios involving strong gravitational fields or high densities, such as in the early universe or within compact astrophysical objects, gravitational forces can significantly influence phase transitions. These influences can arise from pressure variations due to gravity, alterations in the effective potential of a system, or even through the generation of gravitational waves during the phase transition

\section{Results and discussions}\label{Sec4}
\noindent

According to Eq.~\eqref{eq4}, by substituting the equilibrium value of the renormalized order parameter ($\phi(d, T) = \sqrt{-r(d, T)/b}$) into the free energy, the specific heat jump modified by fluctuations is given by the following expression
\begin{equation}
\Delta{C_0}(d, T) = \frac{r^2_0T}{4b_0}\Bigg[\bigg(1 + T_c\frac{\partial{\kappa(d, T)}}{\partial{T}}\bigg)^2 + T^2_c\Big(\tau + \kappa(d, T) \Big)\frac{\partial^2{\kappa(d, T)}}{{\partial{T}}^2}\Bigg].
\label{eq22}
\end{equation}
In the expression of Eq.~\eqref{eq22}, when $\Big(\tau + \kappa(d, T) \Big)\Big\vert_{T = T_{c_d}}$ = 0 vanishes at the renormalized transition point, the quantity $\bigg(1 + T_c\frac{\partial{\kappa(d, T)}}{\partial{T}}\Big\vert_{T = T_{c_d}}\bigg)^2$ provides information about the magnitude of the jump in the specific heat at that same point. If there are significant contributions from scattering states, the height of this jump can either increase or decrease substantially. Additionally, the interaction between $r_0(T - T_c)$ and $\Omega(d, T)$ may result in inhomogeneities in the sample, which can potentially lead to the disappearance of the specific heat jump altogether. Hence, the anomalous part of the specific heat at the renormalized critical point is defined by this relationship.
\begin{equation}
\Delta{C_0}(d, T_{c_{d}}) = \Bigg[\bigg(1 + T_c\frac{\partial{\kappa(d, T)}}{\partial{T}}\Big\vert_{T = T_{c_d}}\bigg)^2 \Bigg]\frac{r^2_0T_{c_d}}{4b_0}.
\label{eq23}
\end{equation}
By setting $\epsilon = \frac{\eta}{r_0T_c}$, we can easily establish the following result
\begin{eqnarray}
\Delta{C_0}(d, T_{c_{d}}) = \left\{  \begin{array}{llll}\vspace{0,3cm}
0, & \textrm{for}\phantom{...} d = 1 \\ \vspace{0,3cm}
(1 + \epsilon)\Delta{C_0}, & \textrm{for} \phantom{...} d = 2 \\ \vspace{0,3cm}
0\phantom{..} \textrm{or} \phantom{..} 4(1 + \epsilon^2)^2\Delta{C_0}, & \textrm{for} \phantom{...} d = 3 \\ \vspace{0,3cm}
(1 - \epsilon)^{-2}\Delta{C_0}, & \textrm{for} \phantom{...} d = 4.
  \end{array}\right.
  \label{eq24}
\end{eqnarray}
where $\Delta{C_0} = \frac{r^2_0T_{c}}{4b_0}$ is the mean-field specific heat jump. The renormalized specific heat jump is influenced by the system's dimension and intrinsic parameters. Self-consistent calculations of Eq.~\eqref{eq9} yield corrections that describe either a disappearance, linear growth, or quasi-exponential growth of the specific heat jump at the renormalized critical point. Linear growth is observed in two-dimensional systems, while exponential growth of the specific heat jump is seen in three-dimensional systems. Notably, one-dimensional systems do not exhibit a specific heat jump, and three-dimensional systems may also lack this jump under certain conditions. In fact, by solving Eq.~\eqref{eq9} for $d$ = 3, we find two solutions; one of these leads to the absence of the specific heat jump at the critical temperature. This phenomenon occurs due to the competition between $r_0(T - T_c)$ and $\Omega(d, T)$, which results in competing phases within the system and, consequently, a competition between the resulting order parameters. Additionally, the disappearance of the specific heat jump in superconducting systems can be attributed to a crossover in the type of order parameters, as discussed in Refs.~\cite{Tanaka}, which leads to a significant loss of entropy at the critical temperature~\cite{Loram,Skocpol}. Exponential growth of the specific heat jump has been observed in families of $\beta$-pyrochlore oxide superconductors~\cite{Hiroi,Shibauchi,Bruhwiler}, which can be explained by a significant lattice contribution coupled with strong fluctuations in the order parameter. However, certain families of high-$T_c$ superconductors, such as TlBaCaCuO and BiSrCaCuO, do not exhibit a conventional specific heat jump at the critical temperature ($T_c$)~\cite{Meingast}. This lack of a clear jump can be attributed to the presence of a partial gap in the electronic density of states at the Fermi level~\cite{Loram2}. Another potential explanation involves the smearing of $T_c$ due to sample inhomogeneities and the competition or interplay between superconductivity and other electronic phases, such as antiferromagnetism. Qualitatively, the results from Eq.~\eqref{eq24} suggest that a small value of $\eta$ can either cancel out or significantly enhance the jump in specific heat at the critical temperature. 

In this review, we provide solutions to the limitations of the standard Ginzburg-Landau theory, specifically focusing on the behaviors of one- and two-dimensional systems. We also explore why certain complex superconducting systems exhibit a disappearance or absence of the specific heat jump. It is important to note that our explanations are tentative and require validation through appropriate experimental studies. The impact of new quadratic coefficients is analyzed in greater detail elsewhere, specifically regarding the non-monotonic variation of the specific heat jump, $\Delta{C_p}/T_{c}$, across the transition for various high-temperature superconductors~\cite{Feulefack}.

\section{Conclusion}\label{Sec5}
\noindent

This paper investigates phase transitions and critical phenomena in materials using the renormalized Ginzburg-Landau theory, along with considerations of symmetry groups and thermal fluctuations. It demonstrates that the thermal responses of a system significantly change as its thickness decreases, presenting challenges for developing a modified Ginzburg-Landau theory that enhances our understanding of the properties of low-dimensional systems. The findings reveal that the behavior of a physical system within the renormalized Ginzburg-Landau theory undergoes multiple phases with qualitatively different characteristics. The study particularly emphasizes the dimensional dependence of thermodynamic parameters and compares these properties with those derived from the standard Ginzburg-Landau theory. We demonstrate that well-controlled fluctuating configurations provide a valuable framework for exploring phase transitions. By applying the renormalized Ginzburg-Landau theory, this work provides theoretical evidence for the existence of new states with unexpected properties, as well as the absence or disappearance of the specific heat jump in certain complex materials, such as under-doped cuprate superconductors.

\section*{Acknowledgments}
\noindent

This work was supported by the Organization for Women in Science for Development ({\color{blue}OWSD}) and the Swedish International Development Cooperation Agency ({\color{blue}Sida}). FEULEFACFACK Ornela Claire would like to especially acknowledge the assistance received from these organizations as part of her Ph.D. studies.

\vspace{1.0cm}

\appendix

\section{Hartree-Fock derivation and treatment of the correction-term $\Omega(d, T)$}\label{App1}
\noindent

To provide a comprehensive understanding, we present the main results related to the self-consistent correction term, denoted as $\Omega(d, T)$. We discussed these approximations in our previous work \cite{Keumo1}, but we did not specify their dimensional expressions. These approximations have been applied in various physical systems \cite{Tsague,Keumo2}.
Indeed, in the continuum limit, the Ginzburg-Landau theory provides a generalized partition function $Z$, which is derived by integrating over all possible configurations $\mathcal{D}\phi(\textbf{r})$ of the order parameter~\cite{Kitazawa}.
\begin{equation}
Z = \int{\mathcal{D}\phi(\textbf{r})} d^{d}\textbf{r}\exp\{-\beta{H_{GL}[\phi]}\}.
%\label{eq13}
\end{equation}
One can then write the (thermal) average value $\langle{A(\phi)}\rangle$ of any thermodynamic parameter $A(\phi)$ as
\begin{equation}
\langle{A(\phi)}\rangle = Z^{-1}\int{\mathcal{D}\phi(\textbf{r})}{A(\phi)} d^{d}\textbf{r}\exp\{-\beta{H_{GL}[\phi]}\}.
%\label{eq14}
\end{equation}
Based on the average value of the order parameter, the susceptibility, which measures the linear response of the system to an external field, is defined as:
\begin{equation}
\langle{\phi}\rangle = \frac{Z^{-1}}{V}\frac{\partial{Z}}{\partial{h}} = \frac{T}{V}\frac{\partial{\ln{Z}}}{\partial{h}} \phantom{..} \Rightarrow \phantom{..} \chi = \frac{\partial\langle{\phi}\rangle}{\partial{h}} = \frac{T}{V}\frac{\partial^2{\ln{Z}}}{\partial{h^2}}.
%\label{eq15}
\end{equation}
More importantly, as important point in our treatment the correlation function relating the order parameter and density at different positions is given by
\begin{equation}
C(\textbf{r}) = \Big\langle\Big\langle{\phi(\textbf{r})\phi(0)}\Big\rangle\Big\rangle = \Big\langle\Big\langle{\Big(\phi(\textbf{r}) - \langle{\phi}\rangle\Big)\Big(\phi(0)-\langle{\phi}\rangle\Big)}\Big\rangle\Big\rangle.
%\label{eq16}
\end{equation}
In this context, for more information on calculating thermodynamic functions, we refer readers to Ref.~\cite{Hohenberg1}. Accordingly, one can introduce the Fourier modes through the decomposition of the spatially dependent order parameter $\phi(\textbf{r})$ into its Fourier component $\phi(\textbf{q})$ and vice-versa, and which are defined as follows:
\begin{eqnarray}
\phi(\textbf{q}) = \frac{1}{V}\int d^{d}\textbf{r}\phi(\textbf{r})\exp\{-i\textbf{q}\cdot\textbf{r}\},\nonumber\\
\phi(\textbf{r}) = \frac{V}{(2\pi)^{d}}\int d^{d}\textbf{q}\phi(\textbf{q})\exp\{i\textbf{q}\cdot\textbf{r}\}. 
%\label{eq17}
\end{eqnarray}

Using the Fourier components and within the framework of the theoretical Landau mean-field theory in the absence of an external field, the corresponding  partition function (normalization) obtained by integration over all possible configurations of the order parameter is obtained as:
\begin{equation}
\mathcal{Z} = \int \mathcal{D}\phi\exp \left[ -\beta\left\lbrace \sum_{q} \big(r_0(T- T_c) + \mathcal{C}q^2\big)\phi(\textbf{q})\phi(-\textbf{q})+\frac{b_{0}}{L^{d}}\sum_{{q_{i}}}\phi(\textbf{q}_{1})\phi(\textbf{q}_{2})\phi{\textbf{q}_{3}}\phi(-\textbf{q}_{1}-\textbf{q}_{2}-\textbf{q}_{3})\right\rbrace \right]
\end{equation}
where $\mathcal{D}\phi(\textbf{q})$ is an integration measure for real modes representing all possible configurations of the order parameter given by
\begin{equation}
\mathcal{D}\phi=\prod_{q}(2\pi)^{-1}d\phi(\textbf{q})d\phi(-\textbf{q}).
\end{equation}
In this respect, $\int \mathcal{D}\phi(\textbf{q})$ defines the functional integral of real modes.
To describe a finite-temperature phase transition, we consider a set of weakly interacting particles or quai-particles. By associating the  $\phi^2$ term with essential fluctuations and linking the $\phi^4$ term to redundant fluctuations, the quartic term acts as an interaction term among the Fourier components of the order parameter. These components represent the fluctuation modes of the order parameter and can be decoupled into a product of two quantities as 
\begin{equation}
\sum_{q_{i}}\phi(\textbf{q}_{1})\phi(\textbf{q}_{2})\phi(\textbf{q}_{3})\phi(-\textbf{q}_{1}-\textbf{q}_{2}-\textbf{q}_{3})\approx 6\sum_{q}\left(\sum_{q'<\Lambda}\langle|\phi(\textbf{q}')\phi(-\textbf{q}')|\rangle \right)\phi(\textbf{q})\phi(-\textbf{q}).
\end{equation}
Thus, the approximation $|\phi(q)|^4 \approx 6\langle|\phi(q)|^2\rangle|\phi(q)|^2$ is used, which assumes that the Fourier components interact only through the mean field produced by other modes.  The factor of 6 here accounts for all possible contractions and $\Lambda$ is a cutoff value corresponding to a lattice periodicity. According to the Hartree-Fock~\cite{Masker,Tucker} approximation, low-dimensional systems are unable to establish long-range order at finite temperatures because of fluctuations in the order parameter. Consequently, fluctuations with wavelengths greater than the coherence length of the sample are considered negligible. This decoupling is justified when the quartic coefficient of the expansion is small.

By making the following replacement:
\begin{equation}
|\phi(q)|^4 \rightarrow 6\langle\phi(q)^2\rangle|\phi(q)|^2,
\end{equation}
in the partition function, the functional integral can be evaluated analytically when dealing with a Gaussian field. By considering the expected fact that the expectation value $\langle\phi(q)^2\rangle$ does not explicitly depend on the wave number $q$, we can subsequently derive according to Eqs. (A1), (A2), (A3), (A4) and (A6) the self-consistent equation for $\Omega(d, T)$
\begin{equation}
\Omega(d, T) = \frac{3b_{0}}{V}\langle\phi(q)^2\rangle = \frac{3b_{0}}{V}\left[ 2\beta\sum_{q}\left( \mathcal{C}q^{2}+r_0(T - T_c)+\Omega(d, T)\right) \right]^{-1}.
\end{equation}
By making a transition to the continuum through the relation 
\begin{equation}
\sum_{q}(...)=L^{d}(2\pi)^{-d}\int_{-\infty}^{+\infty}(...)d^{d}q, 
\end{equation}
we can therefore derive the self-consistent KMDDH equation of the quadratic correction coefficient  $\Omega(d, T)$ given as 
\begin{equation}
 \Omega(d, T) -\eta\left(\frac{T}{T_{c}}\right)\left(\tau +\frac{\Omega(d, T)}{r_{0}T_{c}}\right)^{\frac{d}{2}-1}= 0
\end{equation}
where $\tau = (T - T_{c})/T_{c}$ and $\eta$ denotes the dimension-dependent variational energetic parameter related to the harmonic variance of the order parameter. 
One can therefore make the Gaussian approximation with the following transformation for the quadratic coefficient of the Ginzburg-Landau free energy functional as follows:
\begin{equation}
r_0(T - T_{c}) \rightarrow r_0(T - T_{c}) + \Omega(d, T) 
\end{equation}
 which now takes into account fluctuations and correlations of the order parameter.
These results from the Hartree-Fock approximation show both qualitative and quantitative agreement with specific experimental facts, as described by the Mermin-Wagner-Hohenberg theorem. These results, represented by dashed curves in Fig. 3, pertain specifically to one-dimensional systems and two-dimensional system for large value of $\eta$. The parameter $\eta$ is dimension-dependent and accounts for scattering states contributions~\cite{Keumo1}.

\end{document}